\documentclass[italian,english]{article}
\usepackage[latin9]{inputenc}
\usepackage{color}
\usepackage{array}
\usepackage{longtable}
\usepackage{amssymb}

\providecommand{\tabularnewline}{\\}

\newcommand{\lyxaddress}[1]{
\par {\raggedright #1
\vspace{1.4em}
\noindent\par}
}

\usepackage{babel}

\begin{document}

\title{\textbf{Gravitational wave astronomy: the definitive test for the
{}``Einstein frame versus Jordan frame'' controversy}}

\author{\textbf{Christian Corda}}

\maketitle

\lyxaddress{\begin{center}
Institute for Basic Research, P. O. Box 1577, Palm Harbor, FL 34682,
USA and Associazione Scientifica Galileo Galilei, Via Bruno Buozzi
47, - 59100 PRATO, Italy 
\par\end{center}}

\begin{center}
\textit{E-mail address:} \textcolor{blue}{cordac.galilei@gmail.com}
\par\end{center}
\begin{abstract}
The potential realization of a gravitational wave (GW) astronomy in
next years is a great challenge for the scientific community. By giving
a significant amount of new information, GWs will be a cornerstone
for a better understanding of the universe and of the gravitational
physics. 

In this paper the author shows that the GW astronomy will permit to
solve a captivating issue of gravitation as it will be the definitive
test for the famous {}``Einstein frame versus Jordan frame'' controversy. 

In fact, we show that the motion of the test masses, i.e. the beam
splitter and the mirror in the case of an interferometer, which is
due to the scalar component of a GW, is different in the two frames.
Thus, if a consistent GW astronomy will be realized, an eventual detection
of signals of scalar GWs will permit to discriminate among the two
frames. In this way, a direct evidence from observations will solve
in an ultimate way the famous and long history of the {}``Einstein
frame versus Jordan frame'' controversy.\end{abstract}
\begin{quotation}
\emph{Partially supported by a Research Grant of The R. M. Santilli
Foundations Number RMS-TH-5735A2310}
\end{quotation}

\section{Introduction}

The scientific community hopes in a first direct detection of GWs
in next years \cite{key-1}. The realization of a GW astronomy, by
giving a significant amount of new information, will be a cornerstone
for a better understanding of the universe and of the gravitational
physics. In fact, the discovery of GW emission by the compact binary
system PSR1913+16, composed by two neutron stars \cite{key-2}, has
been, for physicists working in this field, the ultimate thrust allowing
to reach the extremely sophisticated technology needed for investigating
in this field of research. 

In a recent research \cite{key-3}, the author showed that the GW
astronomy will be the definitive test for general relativity, or,
alternatively, a strong endorsement for extended theories of gravity.
In this paper the analysis is improved by showing that, in addition,
the GW astronomy will permit to solve a captivating issue of gravitation
as it will be the ultimate test for the famous {}``Einstein frame
versus Jordan frame'' controversy. 

In fact, the author shows that the motion of test masses, i.e. the
beam splitter and the mirror in the case of an interferometer, in
the field of a scalar GW is different in the two frames. Then, if
a consistent GW astronomy will be realized, an eventual detection
of signals of scalar GWs will permit to discriminate among the two
frames. 

In this way, a direct evidence from observations will solve in an
ultimate way the famous and long history of the {}``Einstein frame
versus Jordan frame'' controversy.

The controversy on conformal frames started from early investigations
\cite{key-4}, till recent analyses \cite{key-5,key-6}, with lots
of effort of famous physicists, see \cite{key-6,key-7,key-8} for
example. In the generalization of the Jordan-Fierz-Brans-Dicke theory
of gravitation \cite{key-9,key-10,key-11}, which is known as scalar-tensor
gravity \cite{key-6,key-12,key-13,key-14}, the gravitational interaction
is mediated by a scalar field together with the usual metric tensor.
Scalar-tensor gravity is present in various frameworks of theoretical
physics, like dilaton gravity in superstring and supergravity theories
\cite{key-15}, like description of braneworld models \cite{key-16},
like conformal equivalents to modified f(R) gravity \cite{key-17},
or in attempts to realize inflation \cite{key-18,key-19,key-20} and
to obtain dark energy \cite{key-21,key-22}. Scalar-tensor gravity
arises from the conviction of lots of scientists that every modern
theoretical attempt to unify gravity with the remaining interactions
requires the introduction of scalar fields \cite{key-12}. An ultimate
endorsement for the viability of scalar-tensor gravity could arrive
from detection of GWs, see \cite{key-3} for details. 

The {}``Einstein frame versus Jordan frame'' controversy started
because some authors claimed that scalar-tensor gravity is unreliable
in the Jordan frame, leading to the problem of negative kinetic energies
\cite{key-23,key-24,key-25}. On the other hand, the Einstein frame
version of scalar-tensor gravity, which is obtained by the conformal
rescaling of the metric \cite{key-26,key-27,key-28,key-29}

\begin{equation}
\tilde{g}_{ab}=\varphi g_{ab}\label{eq: conforme}\end{equation}
and a nonlinear scalar field redefinition \cite{key-26,key-28}

\begin{equation}
\begin{array}{ccc}
d\tilde{\varphi}=\frac{1}{k}\frac{d\varphi}{\varphi} & \Longrightarrow & \tilde{\varphi}=\tilde{\varphi}_{0}+\frac{1}{k}\ln\frac{\varphi}{\varphi_{0}},\end{array}\label{eq: rescaling}\end{equation}
has a positive definite energy \cite{key-27}. In this paper Latin
indices are used for 4-dimensional quantities, Greek indices for 3-dimensional
ones and the author works with $G=1$, $c=1$ and $\hbar=1$ (natural
units). $k$ in Eqs. (\ref{eq: rescaling}) is defined like $k\equiv\sqrt{\frac{16\pi}{|2\omega+3|}}$
and such a notation has not to be confused with other notations in
the literature (in various books and papers $k$ represents the spatial
curvature of Universe, see \cite{key-30} for example). $\varphi$
is the fundamental scalar field of scalar-tensor gravity \cite{key-6,key-12,key-13,key-14},
$\omega$ is the Brans-Dicke parameter \cite{key-11}, $\tilde{\varphi}$
is the \textquotedblleft{}conformal scalar field\textquotedblright{}
\cite{key-26} and $\varphi_{0}$ and $\tilde{\varphi}_{0}$ are constants
that represent the {}``zero values'' of $\varphi$ and $\tilde{\varphi}$. 

In general, analyses in the Einstein frame are simpler concerning
the field equations, but the connection with particle physics is more
difficult than in the Jordan frame. Thus, there are authors who use
the Einstein frame as a mathematical artifice to solve the field equations
and then return in the Jordan frame to compare with astrophysics observations
\cite{key-17,key-22}. Other authors claim that the two conformal
frames are equivalent \cite{key-28}. Others again are not interested
in the problem \cite{key-5}. Different positions of various authors
have been discussed in \cite{key-27} and, at the present time, the
debate remains open \cite{key-5,key-6,key-17,key-22,key-28,key-29}.
The controversy on conformal frames could appear a purely technical
one. Actually, it is very important as the physical predictions of
a classical theory of gravity, or of a dark energy cosmological scenario,
are deeply affected by the choice of the conformal frame. Thus, the
fundamental question is: which is \textit{the physical frame} of observations?
Using of conformal transformations to perform analyses in the Einstein
frame abounds in the literature, with divergence of opinions between
different authors \cite{key-5,key-6,key-17,key-22,key-23,key-24,key-25,key-28,key-29}.
The motion in the Einstein frame \emph{is not geodesic} \cite{key-26},
a key point which strongly endorses deviations from equivalence principle
and non-metric gravity theories in the Einstein frame \cite{key-6,key-26,key-31,key-32}.
Thus, some authors claim that physics must be different in the two
different frames, see \cite{key-31,key-32} for example. Another important
point concerns doubts on the physical equivalence in respect to the
Cauchy problem \cite{key-33,key-34}.

\section{A review of some important issues}

\subsection{Gravitational waves in scalar-tensor gravity: derivation in the Jordan
frame}

In order to better understand the results of this paper it is useful
to sketch the derivation of GWs in scalar-tensor gravity and in the
Jordan frame \cite{key-39}.

The most general action of scalar-tensor theories of gravity in four
dimensions and in the Jordan frame is given by \cite{key-33,key-36}

\begin{equation}
S=\int d^{4}x\sqrt{-g}[f(\phi)R+\frac{1}{2}g^{mn}\phi_{;m}\phi_{;n}-V(\phi)+\mathcal{L}_{(matter)}].\label{eq: scalar-tensor}\end{equation}

Choosing

\begin{equation}
\begin{array}{ccc}
\varphi=f(\phi) & \omega(\varphi)=\frac{f(\phi)}{2f'(\phi)} & W(\varphi)=V(\phi(\varphi))\end{array}\label{eq: scelta}\end{equation}

Eq. (\ref{eq: scalar-tensor}) reads

\begin{equation}
S=\int d^{4}x\sqrt{-g}[\varphi R-\frac{\omega(\varphi)}{\varphi}g^{mn}\varphi_{;m}\varphi_{;n}-W(\varphi)+\mathcal{L}_{(matter)}],\label{eq: scalar-tensor2}\end{equation}

which is a generalization of the Jordan-Fierz-Brans-Dicke theory \cite{key-9,key-10,key-11}.

By varying the action (\ref{eq: scalar-tensor2}) with respect to
$g_{mn}$ and to the scalar field $\varphi$ the field equations are
obtained \cite{key-33,key-36} 

\begin{equation}
\begin{array}{c}
G_{mn}=-\frac{4\pi\tilde{G}}{\varphi}T_{mn}+\frac{\omega(\varphi)}{\varphi^{2}}(\varphi_{;m}\varphi_{;n}-\frac{1}{2}g_{mn}g^{ab}\varphi_{;a}\varphi_{;b})+\\
\\+\frac{1}{\varphi}(\varphi_{;mn}-g_{mn}\square\varphi)+\frac{1}{2\varphi}g_{mn}W(\varphi)\end{array}\label{eq: einstein-general}\end{equation}
with associated a Klein - Gordon equation for the scalar field

\begin{equation}
\square\varphi=\frac{1}{2\omega(\varphi)+3}(-4\pi\tilde{G}T+2W(\varphi)+\varphi W'(\varphi)+\frac{d\omega(\varphi)}{d\varphi}g^{mn}\varphi_{;m}\varphi_{;n}).\label{eq: KG}\end{equation}

In the above equations $T_{mn}$ is the ordinary stress-energy tensor
of the matter and $\tilde{G}$ is a dimensional, strictly positive,
constant. The Newton constant is replaced by the effective coupling

\begin{equation}
G_{eff}=-\frac{1}{2\varphi},\label{eq: newton eff}\end{equation}

which is, in general, different from $G$. General relativity is obtained
when the scalar field coupling is 

\begin{equation}
\varphi=const.=-\frac{1}{2}.\label{eq: varphi}\end{equation}

To study GWs, the linearized theory in vacuum ($T_{mn}=0$) with a
little perturbation of the background has to be analysed \cite{key-30,key-36}.
The background is assumed given by the Minkowskian background plus
$\varphi=\varphi_{0}$ and $\varphi_{0}$ is also assumed to be a
minimum for $W$ \cite{key-36}

\begin{equation}
W\simeq\frac{1}{2}\alpha\delta\varphi^{2}\Rightarrow W'\simeq\alpha\delta\varphi.\label{eq: minimo}\end{equation}

Putting

\begin{equation}
\begin{array}{c}
g_{mn}=\eta_{\mu mn}+h_{mn}\\
\\\varphi=\varphi_{0}+\delta\varphi.\end{array}\label{eq: linearizza}\end{equation}

and, to first order in $h_{mn}$ and $\delta\varphi$, if one calls
$\widetilde{R}_{mnrs}$ , $\widetilde{R}_{mn}$ and $\widetilde{R}$
the linearized quantity which correspond to $R_{mnns}$ , $R_{mn}$
and $R$, the linearized field equations are obtained \cite{key-36}

\begin{equation}
\begin{array}{c}
\widetilde{R}_{mn}-\frac{\widetilde{R}}{2}\eta_{mn}=-\partial_{m}\partial_{n}\Phi+\eta_{mn}\square\Phi\\
\\{}\square\Phi=m^{2}\Phi,\end{array}\label{eq: linearizzate1}\end{equation}

where

\begin{equation}
\begin{array}{c}
\Phi\equiv-\frac{\delta\varphi}{\varphi_{0}}\\
\\m^{2}\equiv\frac{\alpha\varphi_{0}}{2\omega+3}.\end{array}\label{eq: definizione}\end{equation}

The case in which it is $\omega=const.$ and $W=0$ in Eqs. (\ref{eq: einstein-general})
and (\ref{eq: KG}) has been analysed in \cite{key-36} with a treatment
which generalized the {}``canonical'' linearization of general relativity
\cite{key-30}.

For a sake of completeness, let us complete the linearization process
by following \cite{key-36}.

The linearized field equations become

\begin{equation}
\begin{array}{c}
\widetilde{R}_{mn}-\frac{\widetilde{R}}{2}\eta_{mn}=\partial_{m}\partial_{n}\Phi+\eta_{mn}\square\Phi\\
\\\square\Phi=0\end{array}\label{eq: linearizzate2}\end{equation}

Let us put 

\begin{equation}
\begin{array}{c}
\bar{h}_{mn}\equiv h_{mn}-\frac{h}{2}\eta_{mn}+\eta_{mn}\Phi\\
\\\bar{h}\equiv\eta^{mn}\bar{h}_{mn}=-h+4\Phi,\end{array}\label{eq: h barra}\end{equation}

with $h\equiv\eta^{mn}h_{mn}$, where the inverse transform is the
same

\begin{equation}
\begin{array}{c}
h_{mn}=\bar{h}_{mn}-\frac{\bar{h}}{2}\eta_{mn}+\eta_{mn}\Phi\\
\\h=\eta^{mn}h_{mn}=-\bar{h}+4\Phi.\end{array}\label{eq: h}\end{equation}

By putting the first of Eqs. (\ref{eq: h}) in the first of the field
equations (\ref{eq: linearizzate2}) we get

\begin{equation}
\square\bar{h}_{mn}-\partial_{m}(\partial^{a}\bar{h}_{an})-\partial_{n}(\partial^{a}\bar{h}_{an})+\eta_{mn}\partial^{b}(\partial^{a}\bar{h}_{ab}).\label{eq: onda}\end{equation}

Now, let us consider the gauge transform (Lorenz condition)

\begin{equation}
\begin{array}{c}
\bar{h}_{mn}\rightarrow\bar{h}'_{mn}=\bar{h}_{mn}-\partial_{(m}\epsilon_{n)}+\eta_{mn}\partial^{a}\epsilon_{a}\\
\\\bar{h}\rightarrow\bar{h}'=\bar{h}+2\partial^{a}\epsilon_{a}\\
\\\Phi\rightarrow\Phi'=\Phi\end{array}\label{eq: gauge lorenzt}\end{equation}

with the condition $\square\epsilon_{n}=\partial^{m}\bar{h}_{mn}$
for the parameter $\epsilon^{\mu}$. We obtain

\begin{equation}
\partial^{\mu}\bar{h}'_{mn}=0,\label{eq: cond lorentz}\end{equation}

and, omitting the $'$, the field equations can be rewritten like

\begin{equation}
\square\bar{h}_{mn}=0\label{eq: onda T}\end{equation}

\begin{equation}
\square\Phi=0;\label{eq: onda S}\end{equation}

solutions of Eqs. (\ref{eq: onda T}) and (\ref{eq: onda S}) are
plan waves:

\begin{equation}
\bar{h}_{mn}=A_{mn}(\overrightarrow{k})\exp(ik^{a}x_{a})+c.c.\label{eq: sol T}\end{equation}

\begin{equation}
\Phi=a(\overrightarrow{k})\exp(ik^{a}x_{a})+c.c.\label{eq: sol S}\end{equation}

Thus, Eqs. (\ref{eq: onda T}) and (\ref{eq: sol T}) are the equation
and the solution for the tensor waves exactly like in general relativity
\cite{key-30}, while Eqs. (\ref{eq: onda S}) and (\ref{eq: sol S})
are respectively the equation and the solution for the scalar massless
mode \cite{key-36}.

The solutions (\ref{eq: sol T}) and (\ref{eq: sol S}) take the conditions

\begin{equation}
\begin{array}{c}
k^{a}k_{a}=0\\
\\k^{m}A_{mn}=0,\end{array}\label{eq: vincoli}\end{equation}

which arises respectively from the field equations and from Eq. (\ref{eq: cond lorentz}).

The first of Eqs. (\ref{eq: vincoli}) shows that perturbations have
the speed of light, the second the transverse effect of the field.

Fixed the Lorenz gauge, another transformation with $\square\epsilon^{m}=0$
can be made; let us take

\begin{equation}
\begin{array}{c}
\square\epsilon^{m}=0\\
\\\partial_{m}\epsilon^{m}=-\frac{\bar{h}}{2}+\Phi,\end{array}\label{eq: gauge2}\end{equation}

which is permitted because $\square\Phi=0=\square\bar{h}$. We obtain

\begin{equation}
\begin{array}{ccc}
\bar{h}=2\Phi & \Rightarrow & \bar{h}_{mn}=h_{mn}\end{array},\label{eq: h ug h}\end{equation}

i.e. $h_{mn}$ is a transverse plane wave too. The gauge transformations
\cite{key-36}

\begin{equation}
\begin{array}{c}
\square\epsilon^{m}=0\\
\\\partial_{m}\epsilon^{m}=0,\end{array}\label{eq: gauge3}\end{equation}

enable the conditions

\begin{equation}
\begin{array}{c}
\partial^{m}\bar{h}_{mn}=0\\
\\\bar{h}=2\Phi.\end{array}\label{eq: vincoli 2}\end{equation}

Considering a wave propagating in the positive $z$ direction 

\begin{equation}
k^{m}=(k,0,0k),\label{eq: k}\end{equation}

the second of Eqs. (\ref{eq: vincoli}) implies

\begin{equation}
\begin{array}{c}
A_{0\nu}=-A_{3\nu}\\
\\A_{\nu0}=-A_{\nu3}\\
\\A_{00}=-A_{30}+A_{33}.\end{array}\label{eq: A}\end{equation}

Now, let us see the freedom degrees of $A_{mn}$. We were started
with 10 components ($A_{mn}$ is a symmetric tensor); 3 components
have been lost for the transverse condition, more, the condition (\ref{eq: h ug h})
reduces the components to 6. One can take $A_{00}$, $A_{11}$, $A_{22}$,
$A_{21}$, $A_{31}$, $A_{32}$ like independent components; another
gauge freedom can be used to put to zero three more components (i.e.
only three of $\epsilon^{m}$ can be chosen, the fourth component
depends from the others by $\partial_{m}\epsilon^{m}=0$).

Then, by taking 

\begin{equation}
\begin{array}{c}
\epsilon_{m}=\tilde{\epsilon}_{m}(\overrightarrow{k})\exp(ik^{a}x_{a})+c.c.\\
\\k^{m}\tilde{\epsilon}_{m}=0,\end{array}\label{eq: ancora gauge}\end{equation}

the transform law for $A_{mn}$ is (see Eqs. (\ref{eq: gauge lorenzt})
and (\ref{eq: sol T}))

\begin{equation}
A_{mn}\rightarrow A'_{mn}=A_{mn}-2ik(_{m}\tilde{\epsilon}_{n}).\label{eq: trasf. tens.}\end{equation}

Thus, the six components of interest are

\begin{equation}
\begin{array}{ccc}
A_{00} & \rightarrow & A_{00}+2ik\tilde{\epsilon}_{0}\\
A_{11} & \rightarrow & A_{11}\\
A_{22} & \rightarrow & A_{22}\\
A_{21} & \rightarrow & A_{21}\\
A_{31} & \rightarrow & A_{31}-ik\tilde{\epsilon}_{1}\\
A_{32} & \rightarrow & A_{32}-ik\tilde{\epsilon}_{2}.\end{array}\label{eq: sei tensori}\end{equation}

The physical components of $A_{mn}$ are the gauge-invariants $A_{11}$,
$A_{22}$ and $A_{21}$. One can choose $\tilde{\epsilon}_{n}$ to
put equal to zero the others.

The scalar field is obtained by Eq. (\ref{eq: h ug h}):

\begin{equation}
\bar{h}=h=h_{11}+h_{22}=+2\Phi.\label{eq: trovato scalare}\end{equation}

In this way, the total perturbation of a GW propagating in the $z-$
direction in this gauge is

\begin{equation}
h_{\mu\nu}(t+z)=h^{+}(t+z)e_{\mu\nu}^{(+)}+h^{\times}(t+z)e_{\mu\nu}^{(\times)}+\Phi(t+z)e_{\mu\nu}^{(s)}.\label{eq: perturbazione totale}\end{equation}

The term $h^{+}(t+z)e_{\mu\nu}^{(+)}+h^{\times}(t+z)e_{\mu\nu}^{(\times)}$
describes the two standard (i.e. tensor) polarizations of GWs which
arises from general relativity in the TT gauge \cite{key-30}, while
the term $\Phi(t+z)e_{\mu\nu}^{(s)}$ is the extension of the TT gauge
to the scalar-tensor case \cite{key-36}. The correspondent line element
results \cite{key-36}

\begin{equation}
ds^{2}=-dt^{2}+dz^{2}+(1+h^{+}+\Phi)dx^{2}+(1-h^{+}+\Phi)dy^{2}+2h^{\times}dxdy.\label{eq: metrica TT super totale}\end{equation}

This is the case of massless GWs in scalar-tensor gravity. 

By removing the assumptions $\omega=const.$ and $W=0$ in Eqs. (\ref{eq: einstein-general})
and (\ref{eq: KG}) the analysis can be realized for the case of massive
GWs. 

In that case, again $\widetilde{R}_{mnrs}$ and Eqs. (\ref{eq: linearizzate1})
are invariants for gauge transformations \cite{key-35}

\begin{equation}
\begin{array}{c}
h_{mn}\rightarrow h'_{mn}=h_{mn}-\partial_{(m}\epsilon_{n)}\\
\\\Phi\rightarrow\Phi'=\Phi;\end{array}\label{eq: gauge}\end{equation}

then 

\begin{equation}
\bar{h}_{mn}\equiv h_{mn}-\frac{h}{2}\eta_{mn}+\eta_{mn}\Phi\label{eq: ridefiniz}\end{equation}

can be defined, and, by considering the transform for the parameter
$\epsilon^{\mu}$

\begin{equation}
\square\epsilon_{n}=\partial^{m}\bar{h}_{mn},\label{eq:lorentziana}\end{equation}
 a gauge similar to the Lorenz one of electromagnetic waves can be
chosen in this case too

\begin{equation}
\partial^{m}\bar{h}_{mn}=0.\label{eq: cond lorentz 2}\end{equation}
 Thus, the field equations read like

\begin{equation}
\square\bar{h}_{mn}=0\label{onda T 1}\end{equation}

\begin{equation}
\square\Phi=m^{2}\Phi.\label{onda S 1}\end{equation}

Solutions of Eqs. (\ref{onda T 1}) and (\ref{onda S 1}) are plan
waves again

\begin{equation}
\bar{h}_{mn}=A_{mn}(\overrightarrow{p})\exp(ip^{a}x_{a})+c.c.\label{eq: sol T 1}\end{equation}

\begin{equation}
\Phi=a(\overrightarrow{p})\exp(iq^{a}x_{a})+c.c.\label{eq: sol S 1}\end{equation}

where now

\begin{equation}
\begin{array}{ccc}
k^{a}\equiv(\omega,\overrightarrow{p}) &  & \omega=p\equiv|\overrightarrow{p}|\\
\\q^{a}\equiv(\omega_{mass},\overrightarrow{p}) &  & \omega_{mass}=\sqrt{m^{2}+p^{2}}.\end{array}\label{eq: k e q}\end{equation}

Again, in Eqs. (\ref{onda T 1}) and (\ref{eq: sol T 1}) the equation
and the solution for the tensor waves exactly like in general relativity
\cite{key-30} have been obtained, while Eqs. (\ref{onda S 1}) and
(\ref{eq: sol S 1}) are respectively the equation and the solution
for the scalar mode which now is massive \cite{key-35}.

The fact that the dispersion law for the modes of the scalar massive
field $\Phi$ is not linear has to be emphasized. The velocity of
every tensor mode $\bar{h}_{mn}$ is the light speed $c$, but the
dispersion law (the second of Eq. (\ref{eq: k e q})) for the modes
of $\Phi$ is that of a massive field which can be discussed like
a wave-packet \cite{key-35}. Also, the group-velocity of a wave-packet
of $\Phi$ centred in $\overrightarrow{p}$ is \cite{key-35}

\begin{equation}
\overrightarrow{v_{G}}=\frac{\overrightarrow{p}}{\omega_{mass}},\label{eq: velocita' di gruppo}\end{equation}

which is exactly the velocity of a massive particle with mass $m$
and momentum $\overrightarrow{p}$.

From the second of Eqs. (\ref{eq: k e q}) and Eq. (\ref{eq: velocita' di gruppo})
it is simple to obtain:

\begin{equation}
v_{G}=\frac{\sqrt{\omega_{mass}^{2}-m^{2}}}{\omega_{mass}}.\label{eq: velocita' di gruppo 2}\end{equation}

If one wants a constant speed of the wave-packet, it has to be \cite{key-35}

\begin{equation}
m=\sqrt{(1-v_{G}^{2})}\omega_{mass}.\label{eq: relazione massa-frequenza}\end{equation}

Again, the analysis can remain in the Lorenz gauge with transformations
of the type $\square\epsilon_{\nu}=0$; this gauge gives a condition
of transverse effect for the tensor part of the field: $k^{m}A_{mn}=0$,
but it does not give the transverse effect for the total field $h_{mn}$.
From Eq. (\ref{eq: ridefiniz}) we get

\begin{equation}
h_{mn}=\bar{h}_{mn}-\frac{\bar{h}}{2}\eta_{mn}+\eta_{mn}\Phi.\label{eq: ridefiniz 2}\end{equation}

At this point, in the massless case we could put

\begin{equation}
\begin{array}{c}
\square\epsilon^{m}=0\\
\\\partial_{m}\epsilon^{m}=-\frac{\bar{h}}{2}+\Phi,\end{array}\label{gauge massiccia}\end{equation}

which gives the total transverse effect of the field. But in the massive
case this is impossible. In fact, by applying the D' Alembertian operator
to the second of Eqs. (\ref{gauge massiccia}) and by using the field
equations (\ref{onda T 1}) and (\ref{onda S 1}) one obtains

\begin{equation}
\square\epsilon^{m}=+m^{2}\Phi,\label{eq: contrasto}\end{equation}

which is in contrast with the first of Eqs. (\ref{gauge massiccia}).
In the same way, it is possible to show that it does not exist any
linear relation between the tensor field $\bar{h}_{mn}$ and the scalar
field $\Phi$ \cite{key-35}. Thus, a gauge in which $h_{mn}$ is
purely spatial cannot be chosen (i.e. we cannot choose $h_{m0}=0,$
see eq. (\ref{eq: ridefiniz 2})). But the traceless condition to
the field $\bar{h}_{mn}$ can be enabled \cite{key-35}

\begin{equation}
\begin{array}{c}
\square\epsilon^{m}=0\\
\\\partial_{m}\epsilon^{m}=-\frac{\bar{h}}{2}.\end{array}\label{eq: gauge traceless}\end{equation}

These equations imply

\begin{equation}
\partial^{m}\bar{h}_{mn}=0.\label{eq: vincolo}\end{equation}

To enable the conditions $\partial_{m}\bar{h}^{mn}$ and $\bar{h}=0$
transformations like

\begin{equation}
\begin{array}{c}
\square\epsilon^{m}=0\\
\\\partial_{m}\epsilon^{m}=0\end{array}\label{eq: gauge 3}\end{equation}

can be used and, taking $\overrightarrow{p}$ in the $z$ direction,
a gauge in which only $A_{11}$, $A_{22}$, and $A_{12}=A_{21}$ are
different to zero can be chosen. The condition $\bar{h}=0$ gives
$A_{11}=-A_{22}$. Now, by putting these equations in Eq. (\ref{eq: ridefiniz 2})
we obtain

\begin{equation}
h_{mn}(t,z)=h^{+}(t-z)e_{mn}^{(+)}+h^{\times}(t-z)e_{mn}^{(\times)}+\Phi(t-v_{G}z)\eta_{mn}.\label{eq: perturbazione totale 2}\end{equation}

Again, the term $h^{+}(t-z)e_{mn}^{(+)}+h^{\times}(t-z)e_{mn}^{(\times)}$
describes the two standard (i.e. tensor) polarizations of GWs which
arise from general relativity \cite{key-30}, while the term $\Phi(t-v_{G}z)\eta_{mn}$
is the scalar massive field arising from scalar-tensor gravity. In
this case the associated line element results

\begin{equation}
ds^{2}=-(1+\Phi)dt^{2}+(1+\Phi)dz^{2}+(1+h^{+}+\Phi)dx^{2}+(1-h^{+}+\Phi)dy^{2}+2h^{\times}dxdy.\label{eq: metrica TT super totale 2}\end{equation}

\subsection{Quadrupole, dipole and monopole modes}

We emphasize that in this Subsection we closely follow the papers
\cite{key-40,key-41}.

In the framework of GWs, the more important difference between general
relativity and scalar-tensor gravity is the existence, in the latter,
of dipole and monopole radiation \cite{key-40}. In general relativity,
for slowly moving systems, the leading multipole contribution to gravitational
radiation is the quadrupole one, with the result that the dominant
radiation-reaction effects are at order $(\frac{v}{c})^{5}$, where
$v$ is the orbital velocity. The rate, due to quadrupole radiation
in general relativity, at which a binary system loses energy is given
by \cite{key-40}

\begin{equation}
(\frac{dE}{dt})_{quadrupole}=-\frac{8}{15}\eta^{2}\frac{m^{4}}{r^{4}}(12v^{2}-11\dot{r}^{2}).\label{eq:  Will}\end{equation}

$\eta$ and $m$ are the reduced mass parameter and total mass, respectively,
given by $\eta=\frac{m_{1}m_{2}}{(m_{1}+m_{2})^{2}}$ , and $m=m_{1}+m_{2}$
.

$r,$ $v,$ and $\dot{r}$ represent the orbital separation, relative
orbital velocity, and radial velocity, respectively.

In scalar-tensor gravity, Eq. (\ref{eq:  Will}) is modified by corrections
to the coefficients of $O(\frac{1}{\omega})$, where $\omega$ is
the Brans-Dicke parameter (scalar-tensor gravity also predicts monopole
radiation, but in binary systems it contributes only to these $O(\frac{1}{\omega})$
corrections) \cite{key-40}. The important modification in scalar-tensor
gravity is the additional energy loss caused by dipole modes. By analogy
with electrodynamics, dipole radiation is a $(v/c)^{3}$ effect, potentially
much stronger than quadrupole radiation. However, in scalar-tensor
gravity, the gravitational \textquotedblleft{}\emph{dipole moment}\textquotedblright{}
is governed by the difference $s_{1}-s_{2}$ between the bodies, where
$s_{i}$ is a measure of the self-gravitational binding energy per
unit rest mass of each body \cite{key-40}. $s_{i}$ represents the
\textquotedblleft{}\emph{sensitivity}\textquotedblright{} of the total
mass of the body to variations in the background value of the Newton
constant, which, in this theory, is a function of the scalar field
\cite{key-40}: 

\begin{equation}
s_{i}=\left(\frac{\partial(\ln m_{i})}{\partial(\ln G)}\right)_{N}.\label{eq: Will 2}\end{equation}

\emph{$G$} is the effective Newtonian constant at the star and the
subscript $N$ denotes holding baryon number fixed. 

Defining $S\equiv s_{1}-s_{_{2}}$, to first order in $\frac{1}{\omega}$
the energy loss caused by dipole radiation is given by \cite{key-40}
\begin{equation}
(\frac{dE}{dt})_{dipole}=-\frac{2}{3}\eta^{2}\frac{m^{4}}{r^{4}}(12v^{2}-11\dot{r}^{2}).\label{eq:  Will 3}\end{equation}

In scalar-tensor gravity, the sensitivity of a black hole is always
$s_{BH}=0.5$ \cite{key-40}, while the sensitivity of a neutron star
varies with the equation of state and mass. For example, $s_{NS}\approx0.12$
for a neutron star of mass order $1.4M_{\circledcirc}$, being $M_{\circledcirc}$
the solar mass \cite{key-40}. 

Binary black-hole systems are not at all promising for studying dipole
modes because $s_{BH1}-s_{BH2}=0,$ a consequence of the no-hair theorems
for black holes \cite{key-40}. In fact, black holes radiate away
any scalar field, so that a binary black hole system in scalar-tensor
gravity behaves as if general relativity. Similarly, binary neutron
star systems are also not effective testing grounds for dipole radiation
\cite{key-40}. This is because neutron star masses tend to cluster
around the Chandrasekhar limit of $1.4M_{\circledcirc}$, and the
sensitivity of neutron stars is not a strong function of mass for
a given equation of state. Thus, in systems like the binary pulsar,
dipole radiation is naturally suppressed by symmetry, and the bound
achievable cannot compete with those from the solar system \cite{key-40}.
Hence the most promising systems are mixed: BH-NS, BH-WD, or NS-WD. 

The emission of monopole radiation from scalar-tensor gravity is very
important in the collapse of quasi-spherical astrophysical objects
because in this case the energy emitted by quadrupole modes can be
neglected \cite{key-30,key-41}. The authors of \cite{key-41} have
shown that, in the formation of a neutron star, monopole waves interact
with the detectors as well as quadrupole ones. In that case, the field-dependent
coupling strength between matter and the scalar field has been assumed
to be a linear function. In the notation of this paper such a coupling
strength is given by $k^{2}=\frac{16\pi}{|2\omega+3|}$ in Eq. (\ref{eq: rescaling}).
Then \cite{key-41}

\begin{equation}
k^{2}=\alpha_{0}+\beta_{0}(\varphi-\varphi_{0})\label{eq: accoppiamento}\end{equation}

and the amplitude of the scalar polarization results \cite{key-41}

\begin{equation}
\Phi\propto\frac{\alpha_{0}}{d}\label{eq: ampiezza da supernova}\end{equation}

where $d$ is the distance of the collapsing neutron star expressed
in meters.

\subsection{Conformal invariance of the $+$ and $\times$ polarizations}

It is also important to reviewing that the quadrupole modes, i.e.
$+$ and $\times$, are conformal invariants \cite{key-39}.

In standard general relativity the GW-equations in the TT gauge are
\cite{key-30}

\begin{equation}
\square h_{\beta}^{\alpha}=0,\label{eq: dalembert}\end{equation}
where $\square\equiv(-g)^{-1/2}\partial_{a}(-g)^{1/2}g^{ab}\partial_{b}$
is the usual D'Alembert operator. Clearly, matter perturbations do
not appear in (\ref{eq: dalembert}) since scalar and tensor perturbations
do not couple with tensor perturbations in Einstein equations. The
task is now to derive the analogous of Eqs. (\ref{eq: dalembert})
considering the action of scalar-tensor gravity (\ref{eq: scalar-tensor2}).
Matter contributions will be discarded as GWs are analysed in the
linearized theory in vacuum. By following \cite{key-38}, a conformal
analysis helps in this goal. In fact, by considering the conformal
transformation (\ref{eq: conforme}), we obtain the conformal equivalent
Hilbert-Einstein action \begin{equation}
A=\frac{1}{2k}\int d^{4}x\sqrt{-\widetilde{g}}[\widetilde{R}+L(\ln\varphi,(\ln\varphi)_{;a})],\label{eq: conform}\end{equation}

in the Einstein frame, where $L(\ln\varphi,(\ln\varphi)_{;a})$ is
the conformal scalar field contribution derived from \cite{key-38}

\begin{equation}
\tilde{R}_{ab}=R_{ab}+2((\ln\varphi)_{;a}(\ln\varphi)_{;b}-g_{ab}(\ln\varphi)_{;d}(\ln\varphi)^{;d}-\frac{1}{2}g_{ab}(\ln\varphi)^{;d}{}_{;d})\label{eq: conformRicci}\end{equation}

and \begin{equation}
\tilde{R}=\varphi^{-2}+(R-6\square(\ln\varphi)-6(\ln\varphi)_{;d}(\ln\varphi)^{;d}).\label{eq: conformRicciScalar}\end{equation}

In any case, the $L(\ln\varphi,(\ln\varphi)_{;d})$-term does not
affect the GWs-tensor equations, thus it will not be considered any
longer \cite{key-38}.

By starting from the action (\ref{eq: conform}) and deriving the
Einstein-like conformal equations, the GWs equations are \begin{equation}
\widetilde{\square}\widetilde{h}_{\beta}^{\alpha}=0,\label{eq: dalembert conf}\end{equation}

expressed in the conformal metric $\tilde{g}_{ab}.$ As scalar perturbation
does not couple to the tensor part of gravitational waves, it is \cite{key-38}

\begin{equation}
\widetilde{h}_{\beta}^{\alpha}=\widetilde{g}^{\delta\alpha}\delta\widetilde{g}_{\beta\delta}=\varphi^{-2}g^{\delta\alpha}\varphi^{2}\delta g_{\beta\delta}=h_{\beta}^{\alpha},\label{eq: confinvariant}\end{equation}

which means that $h_{\beta}^{\alpha}$ is a conformal invariant.

As a consequence, the plane wave amplitude $h_{\beta}^{\alpha}=h(t)e_{\beta}^{\alpha}\exp(ik_{\beta}x^{\alpha}),$
where $e_{\beta}^{\alpha}$ is the polarization tensor, are the same
in both the Jordan and Einstein frame. The D'Alembert operator transforms
as \cite{key-38}

\begin{equation}
\widetilde{\square}=\varphi^{-2}(\square+2(\ln\varphi)^{;a}\partial_{;a})\label{eq: quadratello}\end{equation}

and this means that the background is changing while the tensor wave
amplitude is fixed.

\section{Geodesic deviation}

The following analysis concerns potential observable effects due by
GWs in order to discriminate \emph{the} \textit{physical frame}. For
this goal, let us use the geodesic deviation equation, which governs
GWs signals in the gauge of the local observer. This gauge is the
locally inertial coordinate system of a laboratory environment on
Earth, where GWs experiments are performed \cite{key-30,key-35,key-36}.
The geodesic deviation equation in the Jordan frame is \cite{key-30}

\begin{equation}
\frac{D^{2}\xi^{d}}{ds^{2}}=\tilde{R}_{abc}^{\quad d}\frac{dx^{c}}{ds}\frac{dx^{b}}{ds}\xi^{a},\label{eq: def geo}\end{equation}

where $\xi^{a}$ is the separation vector between two test masses
\cite{key-30}, i.e. \begin{equation}
\xi^{a}\equiv x_{m1}^{a}-x_{m2}^{a},\label{eq: diff}\end{equation}
 $\frac{D}{ds}$ is the covariant derivative and $s$ the affine parameter
along a geodesic \cite{key-30}. In the Einstein frame the Riemann
tensor rescales as \cite{key-26}

\begin{eqnarray}
R_{abc}^{\quad d} & = & \tilde{R}_{abc}^{\quad d}-2\delta_{[a}^{d}\bigtriangledown_{b]}\bigtriangledown_{c}(\ln\sqrt{\tilde{\varphi}})+\nonumber \\
\nonumber \\ &  & +2g^{de}g_{c[a}\bigtriangledown_{b]}\bigtriangledown_{e}(\ln\sqrt{\tilde{\varphi}})-2\bigtriangledown_{[a}(\ln\sqrt{\tilde{\varphi}})\delta_{b]}^{d}\bigtriangledown_{c}(\ln\sqrt{\tilde{\varphi}})+\label{eq: Riemann c3}\\
\nonumber \\ &  & +2\bigtriangledown_{[a}(\ln\sqrt{\tilde{\varphi}})g_{b]c}g^{de}\bigtriangledown_{e}(\ln\sqrt{\tilde{\varphi}})+2g_{c[a}\delta_{b]}^{d}g^{ef}\bigtriangledown_{e}(\ln\sqrt{\tilde{\varphi}})\bigtriangledown_{f}(\ln\sqrt{\tilde{\varphi}}).\nonumber \end{eqnarray}

Eq. (\ref{eq: Riemann c3}) has to be put into eq. (\ref{eq: def geo}).
Using the contraction properties of $\delta_{b}^{a},$ the symmetry
properties and recalling the normalization condition \cite{key-26,key-30}

\begin{equation}
g_{ac}\frac{dx^{a}}{ds}\frac{dx^{c}}{ds}=1,\label{eq: ricorda}\end{equation}

a bit of algebra gives

\begin{equation}
\frac{D^{2}\xi^{d}}{ds^{2}}=\tilde{R}_{abc}^{\quad d}\frac{dx^{c}}{ds}\frac{dx^{b}}{ds}\xi^{a}+k\frac{D}{ds}(\partial^{d}\tilde{\varphi})\label{eq: geo 3}\end{equation}

Thus, an extra term of the geodesic deviation equations, which is
not present in the Jordan frame, see Eq. (\ref{eq: def geo}), is
present in the Einstein frame, i.e. the term $k\frac{D}{ds}(\partial^{d}\tilde{\varphi}).$

\section{Using gravitational waves to discriminate}

The line element (\ref{eq: metrica TT super totale}) for the scalar
component of massless scalar GWs reduces to 

\begin{equation}
ds^{2}=-dt^{2}+dz^{2}+[1+\Phi(t-z)][dx^{2}+dy^{2}],\label{eq: metrica puramente scalare}\end{equation}

for a wave propagating in the $z$ direction. In the same way the
line element (\ref{eq: metrica TT super totale 2}) for the scalar
component of massive scalar GWs reduces to

\begin{equation}
ds^{2}=[1+\Phi(t-v_{G}z)](-dt^{2}+dz^{2}+dx^{2}+dy^{2}).\label{eq: metrica puramente scalare massiccia}\end{equation}
The cases of massive scalar-tensor gravity and $f(R)$ theories are
totally equivalent \cite{key-3,key-35,key-36,key-37,key-38}. This
is not surprising as it is well known that there is a more general
conformal equivalence between scalar-tensor gravity and $f(R)$ theories
\cite{key-3,key-35,key-36,key-37,key-38}. In fact, $f(R)$ theories
can be conformally reformulated in the Einstein frame by choosing
the conformal rescaling in a slight different way, i.e. $e^{2\tilde{\varphi}}=|f'(R)|$
\cite{key-17,key-38}.

In the Jordan frame the motion of test masses, which is due to scalar
GWs, in the gauge of the local observer is well known \cite{key-35,key-36}.
GWs manifest them-self by exerting tidal forces on the test-masses,
i.e. the mirror and the beam-splitter in the case of an interferometer
\cite{key-35,key-36}. By putting the beam-splitter in the origin
of the coordinate system, the components of the separation vector
are the coordinates of the mirror. At first order in $\Phi$ and $h^{+}$
the total motion of the mirrors due to GWs in massless scalar-tensor
gravity in the Jordan frame is (scalar mode plus quadrupole modes)
\cite{key-35,key-36} \begin{equation}
\delta x_{M}(t)=\frac{1}{2}x_{M0}h^{+}(t)+\frac{1}{2}x_{M0}\Phi(t)\label{eq: spostamento lungo x totale}\end{equation}

and

\begin{equation}
\delta y_{M}(t)=-\frac{1}{2}y_{M0}h^{+}(t)+\frac{1}{2}y_{M0}\Phi(t),\label{eq: spostamento lungo y totale}\end{equation}

where $x_{M0}$ and $y_{M0}$ are the initial (unperturbed) coordinates
of the mirror.

In the case of massive scalar-tensor gravity and of $f(R)$ theories
the total motion of the mirror due to GWs is (scalar mode plus quadrupole
modes) \cite{key-35,key-36}

\begin{equation}
\begin{array}{c}
\delta x_{M}(t)=\frac{1}{2}x_{M0}h^{+}(t)+\frac{1}{2}x_{M0}\Phi(t)\\
\\\delta y_{M}(t)=-\frac{1}{2}y_{M0}h^{+}(t)+\frac{1}{2}y_{M0}\Phi(t)\\
\\\delta z_{M}(t)=-\frac{1}{2}m^{2}z_{M0}\psi(t),\end{array}\label{eq: spostamenti totali}\end{equation}

where \cite{key-35,key-36} \begin{equation}
\ddot{\psi}(t)\equiv\Phi(t).\label{eq: definizione di psi}\end{equation}

Note: the most general definition is $\psi(t-v_{G}z)+a(t-v_{G}z)+b$,
but one assumes only small variations of the positions of the test
masses, thus $a=b=0$ \cite{key-35,key-36}. Then, in the case of
massive GWs a longitudinal component is present because of the presence
of a small mass $m$ \cite{key-35,key-36}. As the interpretation
of $\Phi$ is in terms of a wave-packet, solution of the the Klein
- Gordon equation (\ref{onda S 1}), it is also

\begin{equation}
\psi(t-v_{G}z)=-\frac{1}{\omega^{2}}\Phi(t-v_{G}z).\label{eq: psi 2}\end{equation}

Now, let us see what happens in the Einstein frame. Eqs. (\ref{eq: rescaling})
and (\ref{eq: conforme}) can be used to express the linearized rescaled
scalar field and the linearized conformal transformation. At first
order in $\Phi$ it is 

\begin{equation}
\tilde{\Phi}=\delta\tilde{\varphi}=\frac{1}{k}\frac{\delta\varphi}{\varphi_{0}}=\frac{1}{k}\Phi\label{eq: rescaling scalar}\end{equation}

\begin{equation}
\tilde{g}_{ab}=(1+k\tilde{\Phi})g_{ab}.\label{:eq: conformal}\end{equation}

When the scalar GW passes, it produces an oscillating (linearized)
curvature tensor \cite{key-35,key-36}, plus an addictive component
due to the quantity $k\frac{D}{ds}(\partial^{d}\tilde{\varphi})$
in Eq. (\ref{eq: geo 3}). In the gauge of the local observer all
the correction due to Christoffell-symbols vanish \cite{key-30}.
The gauge of the local observer is a coordinate system that, at first
order in the metric perturbation, moves with the beam splitter and
with its proper reference frame \cite{key-30}. At first order, the
coordinate time $t$ is the same as the proper time in this locally
inertial gauge \cite{key-30}. Hence, putting again the beam-splitter
in the origin of the coordinate system, from Eqs. (\ref{eq: rescaling}),
(\ref{eq: geo 3}) and (\ref{eq: rescaling scalar}) the time evolution
of the coordinates of the mirror in the presence of the scalar GWs,
is \begin{equation}
\frac{d^{2}x_{M}^{\alpha}}{dt^{2}}=\tilde{R}_{0\beta0}^{\quad\alpha}x_{M}^{\beta}+k\frac{\partial^{2}\tilde{\Phi}}{\partial x_{\alpha}\partial x_{\beta}}x_{M}^{\beta}.\label{eq: mostro 2}\end{equation}

In the Einstein frame, using Eq. (\ref{:eq: conformal}), the line
element (\ref{eq: metrica puramente scalare}) for massless GWs rescales
like \begin{equation}
ds^{2}=(1+k\tilde{\Phi})[-dt^{2}+dz^{2}]+(1+2k\tilde{\Phi})[dx^{2}+dy^{2}].\label{eq: metrica puramente scalare conforme}\end{equation}

As it is well known that the linearized Riemann tensor is \emph{gauge
invariant} \cite{key-30}, the components $\tilde{R}_{0\beta0}^{\quad\alpha}x_{M}^{\beta}$
can be computed directly in the gauge of Eq. (\ref{eq: metrica puramente scalare conforme}).
From \cite{key-30} it is:

\begin{equation}
\tilde{R}_{ambn}=\frac{1}{2}\{\partial_{m}\partial_{b}h_{an}+\partial_{n}\partial_{a}h_{mb}-\partial_{a}\partial_{b}h_{mn}-\partial_{m}\partial_{n}h_{ab}\}.\label{eq: riemann lineare}\end{equation}

In the case of eq. (\ref{eq: metrica puramente scalare conforme})
one gets (only the non-zero elements will be explicitly written down)\begin{equation}
\tilde{R}_{010}^{1}=\tilde{R}_{020}^{2}=-k\ddot{\tilde{\Phi}}.\label{eq: riemann lin scalare}\end{equation}

Then, from Eq. (\ref{eq: mostro 2}), the time evolution of the coordinates
of the mirror in the gauge of the local observer is

\begin{equation}
\begin{array}{c}
\ddot{x}_{M}=-k\ddot{\tilde{\Phi}}x_{M}\\
\\\ddot{y}_{M}=-k\ddot{\tilde{\Phi}}y_{M}\\
\\\ddot{z}_{M}=-k\ddot{\tilde{\Phi}}z_{M},\end{array}\label{eq: accelerazione mareale}\end{equation}

i.e., for $j=3$ a third equation is present. Thus, a longitudinal
oscillation, which does not exist in the Jordan frame for massless
scalar GWs, is present in the Einstein frame. By using the perturbation
method \cite{key-30,key-35,key-36} the solutions are: \begin{equation}
\begin{array}{c}
\delta x_{M}(t)=x_{M0}k\tilde{\Phi}(t)\\
\\\delta y_{M}(t)=y_{M0}k\tilde{\Phi}(t)\\
\\\delta z_{M}(t)=z_{M0}k\tilde{\Phi}(t).\end{array}\label{eq: spostamenti 0}\end{equation}

In this way, the longitudinal oscillation makes the total oscillations
of the mirror of the interferometer perfectly isotropic in the Einstein
frame. The third longitudinal oscillation exists as the theory is
non-metric in the Einstein frame.

For a sake of completeness, let us add to Eqs. (\ref{eq: spostamenti 0})
the motion of the mirrors due to the ordinary quadrupole modes \cite{key-39}.
As we have shown in Subsection 2.3 that the quadrupole modes are conformal
invariants, in the Einstein frame the motion of the mirrors due to
quadrupole modes remains unchanged. Hence, we get the total motion:

\begin{equation}
\begin{array}{c}
\delta x_{M}(t)=\frac{1}{2}x_{M0}h^{+}+x_{M0}k\tilde{\Phi}(t)\\
\\\delta y_{M}(t)=-\frac{1}{2}y_{M0}h^{+}+y_{M0}k\tilde{\Phi}(t)\\
\\\delta z_{M}(t)=z_{M0}k\tilde{\Phi}(t).\end{array}\label{eq: spostamenti 0.1}\end{equation}

Now, let us discuss the massive case. Using again eq. (\ref{:eq: conformal}),
at first order in $\tilde{\Phi},$ in the Einstein frame Eq. (\ref{eq: metrica puramente scalare massiccia})
rescales as

\begin{equation}
ds^{2}=(1+2k\tilde{\Phi})(-dt^{2}+dz^{2}+dx^{2}+dy^{2}).\label{eq: metrica puramente scalare massiccia 2}\end{equation}

Taking into account Eq. (\ref{onda S 1}) that, under the transformation
(\ref{eq: rescaling scalar}) remains unaltered, i.e. $\square\tilde{\Phi}=m^{2}\tilde{\Phi},$
and by considering that, from Eqs. (\ref{eq: psi 2}) and (\ref{eq: rescaling scalar})
it is

\begin{equation}
\tilde{\psi}=\frac{1}{k}\psi,\label{eq: psi 3}\end{equation}
Eq. (\ref{eq: riemann lineare}) gives

\begin{equation}
\begin{array}{ccc}
\tilde{R}_{010}^{1}=\tilde{R}_{020}^{2}=-k\ddot{\tilde{\Phi}}, &  & \tilde{R}_{030}^{3}=km\ddot{\tilde{\psi}}.\end{array}\label{eq: componenti riemann again}\end{equation}

To obtain the time evolution of the coordinates of the mirror, one
has to consider the extra term in Eq. (\ref{eq: mostro 2}) too. In
this case, as the scalar field depends from $t-v_{G}z$, at the end
it is 

\begin{equation}
\begin{array}{c}
\ddot{x}_{M}=k\ddot{\tilde{\Phi}}x_{M}\\
\\\ddot{y}_{M}=k\ddot{\tilde{\Phi}}y_{M}\\
\\\ddot{z}_{M}=k(v_{G}^{2}\ddot{\tilde{\Phi}}-m^{2}\ddot{\tilde{\psi}})z_{M},\end{array}\label{eq: accelerazione mareale 3}\end{equation}

Recalling that $m=\sqrt{(1-v_{G}^{2})}\omega$ \cite{key-35,key-36}
and using Eqs. (\ref{eq: psi 2}) and (\ref{eq: psi 3}) the perturbation
method gives the solutions

\begin{equation}
\begin{array}{c}
\delta x_{M}(t)=kx_{M0}\tilde{\Phi}(t)\\
\\\delta y_{M}(t)=ky_{M0}\tilde{\Phi}(t)\\
\\\delta z_{M}(t)=kz_{M0}\tilde{\Phi}(t),\end{array}\label{eq: spostamenti 2}\end{equation}

which are exactly the same of the massless case (\ref{eq: spostamenti 0}).
In fact, even if the non-metric longitudinal motion is different with
respect to the massless case, in the massive case there is also a
metric longitudinal motion. Thus, the sum of the non-metric longitudinal
motion and of the metric longitudinal motion in the massive case results
equal to the total non-metric longitudinal motion in the massless
case. In the massless case the longitudinal motion is totally non-metric.
However, even if the motion of the mirror is the same for massless
and massive scalar GWs in the Einstein frame, in principle, careful
analyses of coincidences between various detectors could permit to
discriminate between massless and massive cases because in the massless
case the speed of the GW is exactly the speed of light, while in the
massive case the speed of the GW is the group velocity $v_{G},$ lower
than the speed of light.

Again, let us add to Eqs. (\ref{eq: spostamenti 2}) the motion of
the mirrors due to the ordinary quadrupole modes \cite{key-39}. We
obtain the total motion \begin{equation}
\begin{array}{c}
\delta x_{M}(t)=\frac{1}{2}x_{M0}h^{+}+kx_{M0}\tilde{\Phi}(t)\\
\\\delta y_{M}(t)=-\frac{1}{2}y_{M0}h^{+}+ky_{M0}\tilde{\Phi}(t)\\
\\\delta z_{M}(t)=kz_{M0}\tilde{\Phi}(t).\end{array}\label{eq: spostamenti 2.1}\end{equation}

Now, let us explain why we are claiming that the GW astronomy will
be the definitive test for the {}``Einstein frame versus Jordan frame''
controversy. In principle, if advanced projects on the detection of
GWs will improve their sensitivity allowing to perform a GW astronomy,
one will only have to look which is the motion of the mirror in respect
to the beam splitter of an interferometer in the locally inertial
coordinate system in order to understand which is the physical frame
of observations. If such a motion will be governed by Eqs. (\ref{eq: spostamento lungo x totale})
and (\ref{eq: spostamento lungo y totale}) for massless scalar waves
or by Eqs. (\ref{eq: spostamenti totali}) for massive scalar waves,
one will conclude that the physical frame of observations is the Jordan
frame. If the motion of the mirror is governed by Eqs. (\ref{eq: spostamenti 0.1})
for massless scalar GWs which are equal to Eqs. (\ref{eq: spostamenti 2.1})
for massive scalar GWs one will conclude that the physical frame of
observations is the Einstein frame. 

On the other hand, such signals will be quite weak. Thus, in order
for the analysis to be useful in practice, we have to provide a specific
application of the proposed method \cite{key-39}. In particular,
we have to compare the trajectories in both of the frames and determine
the experimental sensitivity required to distinguish them. We have
also to compare with the sensitivities of ongoing and future experiments
\cite{key-39}. To make this, we consider an astrophysical event which
produces GWs and which can, in principle, help to simplify the problem.
In Subsection 2.2 we discussed two potential sources of potential
detectable scalar radiation:
\begin{enumerate}
\item mixed binary systems like BH-NS, BH-WD, or NS-WD;
\item the gravitational collapse of quasi-spherical astrophysical objects.
\end{enumerate}
The second source looks propitious because in such a case the energy
emitted by quadrupole modes can be neglected \cite{key-41} (in the
sense that the monopole modes largely exceed the quadrupole ones.
In fact, if the collapse is completely spherical, the quadrupole modes
are totally removed \cite{key-30}). In that case, only the motion
of the test masses due to the scalar component has to be analysed.
Hence, the motion of the test masses in the Jordan frame is given
by

\begin{equation}
\delta x_{M}(t)=\frac{1}{2}x_{M0}\Phi(t)\label{eq: spostamento lungo x}\end{equation}

and

\begin{equation}
\delta y_{M}(t)=\frac{1}{2}y_{M0}\Phi(t),\label{eq: spostamento lungo y}\end{equation}
for massless GWs and by

\begin{equation}
\begin{array}{c}
\delta x_{M}(t)=\frac{1}{2}x_{M0}\Phi(t)\\
\\\delta y_{M}(t)=\frac{1}{2}y_{M0}\Phi(t)\\
\\\delta z_{M}(t)=-\frac{1}{2}m^{2}z_{M0}\psi(t),\end{array}\label{eq: spostamenti}\end{equation}

for massive GWs, while Eqs. (\ref{eq: spostamenti 0}) for massless
GWs and Eqs. (\ref{eq: spostamenti 2}) for massive GWs govern the
motion of the test masses in the Einstein frame. Thus, the problem
is simpler. The authors of \cite{key-41} analysed the interesting
case of the formation of a neutron star through a gravitational collapse.
In that case, they found that a collapse occurring closer than 10
kpc from us (half of our Galaxy) needs a sensitivity of  $3*10^{-23}\mbox{ }\sqrt{Hz}$
at $800\mbox{ }Hz$ (which is the characteristic frequency of such
events) to potential detect the strain which is generated by the scalar
component in the arms of LIGO. 

At the present time, the sensitivity of LIGO at about $800\mbox{ }Hz$
is $10^{-22}\mbox{ }\sqrt{Hz}$ while the sensitivity of the Enhanced
LIGO Goal is predicted to be $8*10^{-22}\mbox{ }\sqrt{Hz}$ at $800\mbox{ }Hz$
\cite{key-1}. Then, for a potential realization of the test proposed
in this paper, we have to hope in Advanced LIGO Baseline High Frequency
and/or in Advanced LIGO Baseline Broadband. In fact, the sensitivity
of these two advanced configuration is predicted to be $6*10^{-23}\mbox{ }\sqrt{Hz}$
at $800\mbox{ }Hz$ \cite{key-1}. If such a sensitivity will be really
achieved, it will be possible to distinguish the different trajectories
of the mirror in the two frames. 

For a sake of completeness \cite{key-39}, we recall that in the case
of standard general relativity the scalar mode is not present. In
that case, the motion of test masses is governed by \cite{key-30}

\begin{equation}
\delta x_{M}(t)=\frac{1}{2}x_{M0}h^{+}(t)\label{eq: spostamento lungo x GR}\end{equation}

and

\begin{equation}
\delta y_{M}(t)=-\frac{1}{2}y_{M0}h^{+}(t).\label{eq: spostamento lungo y GR}\end{equation}

In the case of scalar-tensor gravity, it will be very important to
understand if a longitudinal component will be present. Such a longitudinal
component will be fundamental in order to discriminate between the
two frames. If it will be absent and the motion of the mirror will
be governed by the transverse eqs. (\ref{eq: spostamento lungo x})
and (\ref{eq: spostamento lungo y}) we will conclude that we are
in presence of massless scalar GWs and the physical frame is the Jordan
frame. On the other hand, if it will be present we have two possibility.
If it will be perfectly isotropic with respect the two transverse
oscillations, i.e. the motion of the mirror will be governed by Eqs.
(\ref{eq: spostamenti 0}) or Eqs. (\ref{eq: spostamenti 2}), we
will conclude that the physical frame is the Einstein frame. If it
will not be perfectly isotropic with respect the two transverse oscillations,
i.e. the motion of the mirror will be governed by Eqs. (\ref{eq: spostamenti}),
we will conclude that we are in presence of massive scalar GWs and
the physical frame is the Jordan frame. 

Let us resume the situation by including a Table with 5 rows and 3
columns \cite{key-39}. In the first column we include the 5 models
to be distinuished (general relativity, massless-Jordan, massive-Jordan,
massless-Einstein, massive-Einstein), in the second column we include
the corresponding motion of the mirror and in the third column the
polarizations and the corresponding symmetry properties of the trajectories
\cite{key-39}. 

\begin{longtable}{|>{\raggedright}m{0.2\columnwidth}|>{\raggedright}p{0.5\columnwidth}|>{\raggedright}m{0.5\columnwidth}|}
\hline 
general relativity & $\begin{array}{c}
\delta x_{M}(t)=\frac{1}{2}x_{M0}h^{+}(t)\\
\\\delta y_{M}(t)=-\frac{1}{2}y_{M0}h^{+}(t)\end{array}$ & transverse motion, only $h^{+}$ polarization\tabularnewline
\hline
\hline 
massless-Jordan & $\begin{array}{c}
\delta x_{M}(t)=\frac{1}{2}x_{M0}h^{+}(t)+\frac{1}{2}x_{M0}\Phi(t)\\
\\\delta y_{M}(t)=-\frac{1}{2}y_{M0}h^{+}(t)+\frac{1}{2}y_{M0}\Phi(t)\end{array}$ & transverse motion, $h^{+}$ polarization and $\Phi$ polarization\tabularnewline
\hline 
massive-Jordan & $\begin{array}{c}
\delta x_{M}(t)=\frac{1}{2}x_{M0}h^{+}(t)+\frac{1}{2}x_{M0}\Phi(t)\\
\\\delta y_{M}(t)=-\frac{1}{2}y_{M0}h^{+}(t)+\frac{1}{2}y_{M0}\Phi(t)\\
\\\delta z_{M}(t)=-\frac{1}{2}m^{2}z_{M0}\psi(t)\end{array}$ & transverse and longitudinal motion, $h^{+}$ polarization and $\Phi$
polarization, no-isotropy between transverse and longitudinal motion
due to the scalar component\tabularnewline
\hline 
massless-Einstein & $\begin{array}{c}
\delta x_{M}(t)=\frac{1}{2}x_{M0}h^{+}+kx_{M0}\tilde{\Phi}(t)\\
\\\delta y_{M}(t)=-\frac{1}{2}y_{M0}h^{+}+ky_{M0}\tilde{\Phi}(t)\\
\\\delta z_{M}(t)=kz_{M0}\tilde{\Phi}(t)\end{array}$ & transverse and longitudinal motion, $h^{+}$ polarization and $\Phi$
polarization, the oscillations due to the scalar component are perfectly
isotropic\tabularnewline
\hline 
massive-Einstein & $\begin{array}{c}
\delta x_{M}(t)=\frac{1}{2}x_{M0}h^{+}+kx_{M0}\tilde{\Phi}(t)\\
\\\delta y_{M}(t)=-\frac{1}{2}y_{M0}h^{+}+ky_{M0}\tilde{\Phi}(t)\\
\\\delta z_{M}(t)=kz_{M0}\tilde{\Phi}(t)\end{array}$ & transverse and longitudinal motion, $h^{+}$ polarization and $\Phi$
polarization, the oscillations due to the scalar component are perfectly
isotropic\tabularnewline
\hline
\end{longtable}

Clearly, this is a simple analysis which could be improved by the
realization of a consistent GW astronomy that, by using coincidences
between various detectors and by further improving the sensitivity
of the detectors, could, in principle, enable a better analysis of
the signals that we have discussed.

\section{Conclusion remarks}

Resuming, in this paper we have shown that the GW astronomy will permit
to solve a captivating issue of gravitation, i.e. it will be the definitive
test for the famous {}``Einstein frame versus Jordan frame'' controversy.
In fact, the author has shown that the motion of test masses in the
field of a scalar GW is different in the two frames, thus, if a consistent
GW astronomy will be realized, an eventual detection of scalar GWs
will permit to discriminate among the two frames.

In this way, direct evidences from observations will solve in an ultimate
way the famous and long history of the {}``Einstein frame versus
Jordan frame'' controversy.

\section{Acknowledgements }

The Associazione Scientifica Galileo Galilei has to be thanked for
supporting this paper. I thank an unknown referee for precious advices
which permitted to improve this paper.


\begin{thebibliography}{41}
\bibitem[1]{key-1}The LIGO Scientific Collaboration, Class. Quant.
Grav. 26, 114013 (2009)

\bibitem[2]{key-2}R. A. Hulse and J.H. Taylor - Astrophys. J. Lett.
195, 151 (1975) 

\bibitem[3]{key-3}C. Corda - Int. Journ. Mod. Phys. D, 18, 14, 2275-2282
(2009, Honorable Mention Gravity Research Foundation)

\bibitem[4]{key-4}R. H. Dicke - Phys. Rev. 125, 2163 (1962) 

\bibitem[5]{key-5}M. Roshan, F. Shojai - Phys. Rev. D 80, 043508
(2009)

\bibitem[6]{key-6}C. Will - Liv. Rev. Rel. 4, 4 (2001) updated at
Publication URI: http://www.livingreviews.org/lrr-2006-3 (2006)

\bibitem[7]{key-7}J. D. Barrow and K. Maeda - Nucl. Phys. B 341,
294 (1990) 

\bibitem[8]{key-8}J. P. Mimoso and D. Wands - Phys. Rev. D 51, 477
(1995) 

\bibitem[9]{key-9}P. Jordan - Naturwiss. 26, 417 (1938) 

\bibitem[10]{key-10}M. Fierz - Helv. Phys. Acta 29, 128 (1956) 

\bibitem[11]{key-11}C. Brans and R. H. Dicke - Phys. Rev. 124, 925
(1961) 

\bibitem[12]{key-12}C. M. Will - \emph{Theory and experiment in gravitational
physics} - Cambridge University Press, Cambridge (1993) 

\bibitem[13]{key-13}R. V. Wagoner - Phys. Rev. D 1, 3209 (1970) 

\bibitem[14]{key-14}Y. Fujii and K. Maeda - \textit{The scalar-tensor
theory of gravitation} Cambridge University Press, Cambridge (2003) 

\bibitem[15]{key-15}\foreignlanguage{italian}{M.B. Green, J. Schwarz
and E. Witten\textit{ - Superstring Theory} - Cambridge University
Press, Cambridge (1987)} 

\bibitem[16]{key-16}J. Garriga and T. Tanaka - Phys. Rev. Lett. 84,
2778 (2000) 

\selectlanguage{italian}%
\bibitem[17]{key-17}\foreignlanguage{english}{G. Cognola, E. Elizalde,
S. Nojiri, S.D. Odintsov, L. Sebastiani, S. Zerbini - Phys. Rev. D
77, 046009 (2008)} 

\selectlanguage{english}%
\bibitem[18]{key-18}\foreignlanguage{italian}{A. Guth - Phys.} Rev.\foreignlanguage{italian}{
\textbf{23} 347 (1981)}

\bibitem[19]{key-19}\foreignlanguage{italian}{D. H. Lyth and A. R.
Liddle - \emph{Primordial Density Perturbation}, Cambridge University
Press (2009)} 

\bibitem[20]{key-20}\foreignlanguage{italian}{G. Watson - \emph{An
exposition on inflationary cosmology} - North Carolina University
Press (2000)} 

\bibitem[21]{key-21}G. Esposito-Farese, D. Polarski, and A. A. Starobinsky
- Phys. Rev. Lett. 85, 2236 (2000) 

\bibitem[22]{key-22}E. Elizalde, S. Nojiri, and S. D. Odintsov -
Phys. Rev. D 70, 043539 (2004) 

\bibitem[23]{key-23}P. Teyssandier and P. Tourrenc - J. Math. Phys.
24 2793 (1983) 

\bibitem[24]{key-24}L. Sokolowski - Class. Quant. Grav. 6, 59, 2045
(1989) 

\bibitem[25]{key-25}D.I. Santiago and A. S. Silbergleit - Gen. Rel.
Grav. 32 565 (2000) 

\bibitem[26]{key-26}R. M. Wald - \textit{General Relativity -} The
Universiy Chicago Press, Chicago (1984) 

\bibitem[27]{key-27}G. Magnano and L. M. Sokolowski - Phys. Rev.
D 50, 5039, (1994) 

\bibitem[28]{key-28}N. Deruelle and M. Sasaki, to appear in the Proceedings:
Cosmology, Quantum Vacuum and Zeta Functions, 8-10 March, 2010, pre-print
in arXiv:1007.3563 (2010)

\bibitem[29]{key-29}M. D. Klimek - Class. Quant. Grav. 26, 065005
(2009) 

\bibitem[30]{key-30}\foreignlanguage{italian}{Misner CW, Thorne KS
and Wheeler JA - \emph{Gravitation} - W.H.Feeman and Company - 1973} 

\bibitem[31]{key-31}Y.M. Cho - Phys. Rev. Lett. 68, 3133 (1992)

\bibitem[32]{key-32}S. Capozziello, P. Martin-Moruno, C. Rubano,
Phys. Lett. B 689, 117-121 (2010)

\bibitem[33]{key-33}T. Damour and G. Esposito-Farese - Class. Quant.
Grav. 9, 2093 (1992) 

\bibitem[34]{key-34}M. Salgado - Class. Quant. Grav. 23, 4719 (2006)

\bibitem[35]{key-35}C. Corda - J. Cosmol. Astropart. Phys. JCAP04009
(2007)

\bibitem[36]{key-36}Capozziello S and C. Corda - Int. J. Mod. Phys.
D \textbf{15,} 1119 -1150 (2006)

\bibitem[37]{key-37}S. Capozziello, C. Corda and M. F. De Laurentis
- Phys. Lett. B, 669, 255-259 (2008) 

\bibitem[38]{key-38}S. Capozziello, C. Corda and M. F. De Laurentis
- Mod. Phys. Lett. A 22, 15, 1097-1104 (2007) 

\bibitem[39]{key-39}Private communication with a referee

\bibitem[40]{key-40}Paul D. Scharre and Clifford M. Will, Phys.Rev.
D 65, 042002 (2002)

\bibitem[41]{key-41}J. Novak, and J.M. Ibanez, Astrophys. J. 533,
392-405 (2000)
\end{thebibliography}
\end{document}